\documentclass[fleqn,10pt]{wlscirep}
\usepackage[utf8]{inputenc}
\usepackage[T1]{fontenc}
\usepackage{chemformula}
\usepackage[normalem]{ulem}
\usepackage{rcs}
\usepackage{color,graphicx}
\usepackage{dcolumn}
\usepackage{bm}
\usepackage{chngcntr}
\usepackage{amsmath,amssymb}
\usepackage{doi}
\usepackage{hyperref}
\usepackage{ragged2e}

\usepackage{comment}
\usepackage{tabularx}
\usepackage{makecell}
\usepackage{varwidth}
\usepackage{makecell}
\usepackage{wasysym}
\usepackage{array}
\usepackage{multirow}
\usepackage{soul}
\usepackage{transparent}

\newcommand{\CRA}{CeRh$_{2}$As$_{2}$}

\newcommand{\Tc}{$T_{\textrm{c}}$}
\newcommand{\To}{$T_{\textrm{0}}$\,}

\definecolor{faintblue}{RGB}{0,0,255}
\colorlet{faintblue50}{faintblue!50}

\title{Investigation of \CRA\ order parameters via ultrasound propagation anomalies}
\author[1,2,3,4$\dagger$*]{S. Galeski}
\author[5 $\dagger$]{C. Lee}
\author[3, 8]{F. Bärtl}
\author[3]{J. Sourd}
\author[3]{S. Zherlitsyn}
\author[1]{A.T.M. Breugelmans}
\author[1]{R. Amdouni}
\author[3, 6]{P. Khanenko}
\author[6,7]{E. Hassinger}
\author[6]{S. Khim}
\author[3,8]{J. Wosnitza}
\author[6]{P. Thalmeier}
\author[5]{P. M. R. Brydon}
\author[6,**]{M. Brando}
\affil[1]{
HFML-FELIX, Toernooiveld 7, 6525ED Nijmegen, the Netherlands}

\affil[2]{Institute for Molecules and Materials, Radboud University, Heyendaalseweg 135, 6525 AJ Nijmegen, The Netherlands   }

\affil[3]{Hochfeld-Magnetlabor Dresden (HLD-EMFL) and Würzburg-Dresden Cluster of Excellence ctd.qmat, Helmholtz-Zentrum Dresden-Rossendorf, 01328 Dresden, Germany}
\affil[4]{Physikalisches Institut, Universität Bonn, Nussallee 12,53115 Bonn, Germany}
\affil[5]{Department of Physics and MacDiarmid Institute for Advanced Materials and Nanotechnology, University of Otago, Dunedin, New Zealand}
\affil[6]{Max Planck Institute for Chemical Physics of Solids, Nöthnitzer Stra{\ss}e 40, 01187 Dresden, Germany}
\affil[7]{Institute for Quantum Materials and Technologies, Karlsruhe Institute of Technology, Kaiserstraße 12, 76131 Karlsruhe, Germany}
\affil[8]{Technical University Dresden, Institute for Solid State and Materials Physics, 01062 Dresden, Germany}
\affil[$\dagger$]{contributed equally}
\affil[*]{sgaleski@science.ru.nl}
\affil[**]{brando@cpfs.mpg.de }
\begin{abstract}
Unconventional superconductors with nearly degenerate pairing states are rare. \CRA\ has recently emerged as one of the few existing multi-phase superconductors. It exhibits a first-order phase transition between two distinct superconducting states when a magnetic field is applied along the crystallographic $c$-axis. While this behavior has been linked to locally broken inversion symmetry, a phase diagram based on a multicomponent superconducting order parameter remains a possibility. Furthermore, superconductivity appears to coexist with an ordered state (phase I). Despite being the subject of many studies, little is known about the nature of the order parameters in both superconducting phases and phase I. Here, we use ultrasound-propagation measurements at low temperatures, in high magnetic fields and under hydrostatic pressure to address this issue. Our results strongly suggest that the superconducting order parameter in both phases is single-component, corroborating the role of local non-centrosymmetry in the development of multi-phase superconductivity in \CRA. In addition, analysis of the elastic anomalies within the Landau framework of phase transitions strongly suggests the presence of an incommensurate magnetic order parameter in phase I.\\

\end{abstract}
\begin{document}
\flushbottom
\maketitle
Unconventional superconductivity is one of the most puzzling phenomena in contemporary condensed-matter physics. Unlike conventional Bardeen–Cooper–Schrieffer (BCS) superconductors~\cite{tinkham1996}, in which electrons form Cooper pairs in the most symmetric form (an $s$-wave spin-singlet state) due to phonon-mediated interactions, the superconducting condensate in unconventional superconductors is believed to form through purely electronic or magnetic interactions~\cite{sigrist2005,monthoux2007}. This results in order parameters (OP) of different symmetries ($s, p , d, f...$,)~\cite{sigrist2005}. 
Despite the wide range of superconducting (SC) order parameters, materials in which multiple distinct SC phases, which are directly connected in the phase diagram, have been observed are extremely rare. The only known compounds are the heavy-fermion systems UPt$_{3}$~\cite{fisher1989,hasselbach1989}, U$_{1-x}$Th$_{x}$Be$_{13}$~\cite{heffner1989}, UTe$_{2}$~\cite{braithwaite2019} and YbRh$_{2}$Si$_{2}$~\cite{nguyen2021,levitin2025}. 
Of these materials, only the phase diagram of UPt$_3$ is well-understood. It is widely accepted that the multiple superconducting phases arise from a two-component order parameter where the degeneracy between the two components is lifted by a small symmetry-breaking field~\cite{joynt2002}. This scenario requires materials whose crystal symmetry supports multi-dimensional irreducible representations (IRs), thereby excluding orthorhombic compounds like UTe$_2$.

In 2012, Yoshida~\textit{et al.} proposed a different route towards multi-phase superconductivity that does not require a multi-component OP. They considered a two-dimensional bi-layer electron system with a small interlayer hopping, and an opposite sign of the Rashba spin-orbit coupling (SOC) within the two layers; the latter reflects the `locally' broken inversion symmetry in each layer. When the interlayer coupling is sufficiently weak compared to the SOC, a magnetic field applied perpendicular to the layers leads to a first-order phase transition between an even-parity superconducting (SC) state at low field and an odd-parity SC state at high field~\cite{fisher2023}. However, although superconducting materials with locally broken inversion symmetry are not rare, no such superconductor showing multiple SC phases was found until the discovery of \CRA\ ($Tc$\ = 0.31\,K) in 2021~\cite{khim2021}.

\begin{figure}[ht!]
	\centering
	\includegraphics[width=0.9\textwidth]{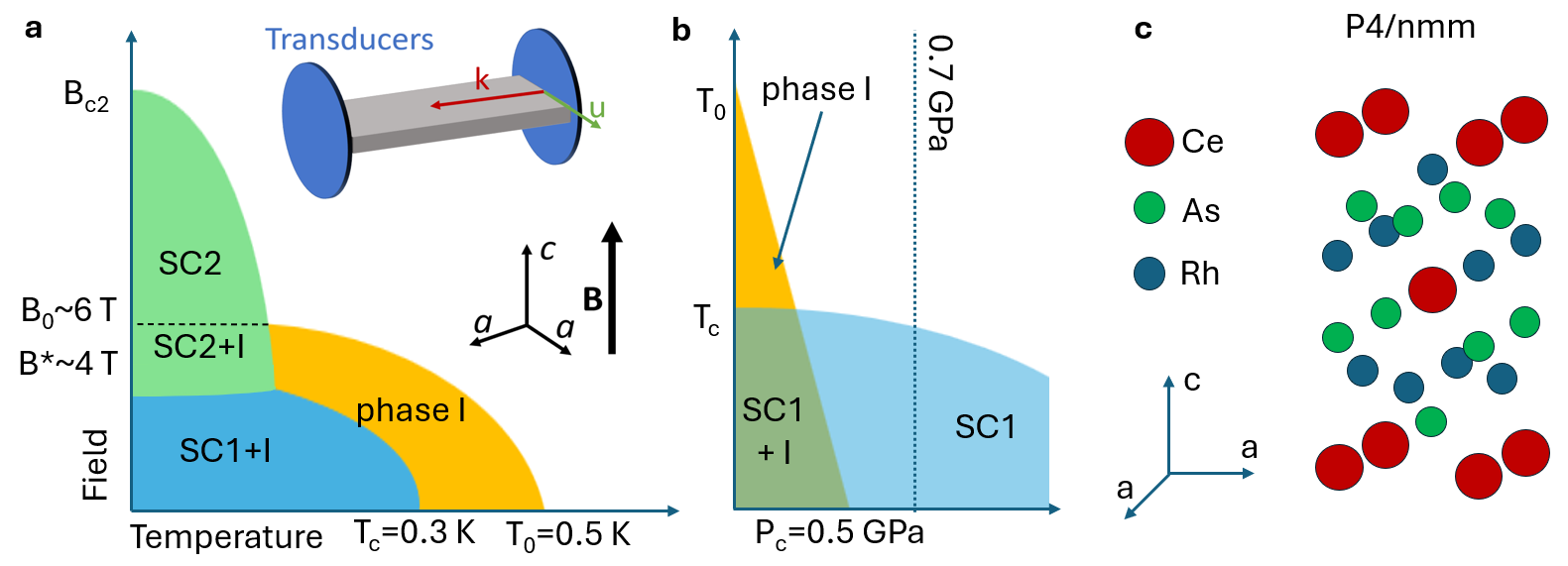}
	\caption{\textbf{Phase diagram of \CRA, crystal structure, and experimental set-up:} a) Schematic phase diagram of \ch{CeRh2As2} for magnetic fields applied along the crystallographic \textit{c}-axis, according to Ref.~\cite{khanenko2025}. Phases SC1 and SC2 represent the superconducting phases found at low temperatures, phase I is the magnetic phase. The inset shows a sketch of the experimental setup used in ultrasound experiments: LiNb transducers attached to the sample produce sound waves, which travel in the direction \textbf{k} with polarization \textbf{u}. b) Schematic phase diagram of \ch{CeRh2As2} under applied pressure at $B = 0$, according to Ref.~\cite{pfeiffer2024}. The dashed line represents the pressure at which ultrasonic experiments in this work were conducted. c) Locally non-centrosymmetric crystal structure of \ch{CeRh2As2} with a body-centered-tetragonal (bct) Ce sublattice. } 
	\label{fig1}
\end{figure}

Indeed, \CRA\ possesses all the ingredients postulated in the theoretical prediction~\cite{yoshida2012,khim2021}: local inversion symmetry breaking in the structure at the Ce sites (see Fig.~\ref{fig1}c), enabling an alternating Rashba SOC between the Ce layers and a quasi-two-dimensional electronic structure in one heavy sheet of the Fermi surface close to the edge of the Brillouin zone, where the SOC is strongest~\cite{hafner2022,cavanagh2022}. The experimentally determined phase diagrams of \CRA\ for magnetic fields applied parallel to and perpendicular to the Ce layers closely resemble those proposed by Yoshida~\textit{et al.}~\cite{khim2021,landaeta2022a,mishra2022,semeniuk2023,chajewski2024,khanenko2025}. In particular, two superconducting phases labelled SC1 and SC2 are observed for $B \parallel c$ and are separated by a weakly first-order phase transition line $B^{*} \approx 4$\,T (cf. Fig.~\ref{fig1}a). However, despite all the experimental results confirming the model proposed by Yoshida~\textit{et al.}, the possibility that the phase diagrams arise from a two-component order parameter, as in UPt$_3$, has not yet been ruled out.

The relevance of the scenario proposed by Yoshida~\textit{et al.} for \CRA\ is complicated by the presence of another ordered state (phase I), which is realized below $T_0 \approx 0.5$\,K~\cite{khim2021}. Phase I was originally proposed as a unique quadrupolar-density-wave state, induced by the quasi-quartet ground state of the crystalline electric field (CEF) of Ce$^{3+}$~\cite{hafner2022,christovam2023,takimoto2008}. More recently, $\mu$SR~\cite{khim2025}, NQR/NMR~\cite{kibune2022,ogata2023,ogata2024} and micro-Hall-probe magnetometry~\cite{juraszek2025} experiments have shown that static local antiferromagnetic (AFM) order occurs below \To\ and coexists microscopically with superconductivity below \Tc, as shown schematically in Fig.~\ref{fig1}. A recent thermodynamic study has also shown that phase I interacts only weakly with superconductivity, which is consistent with the coexistence of the AFM and SC order parameters~\cite{semeniuk2023,khanenko2025}. More importantly, pressure experiments clearly demonstrate that the phase I vanishes at a quantum critical point (QCP) at a pressure of 0.5\,GPa (see Fig.~\ref{fig1}b), whereas superconductivity persists up to a pressure of at least 2.8\,GPa~\cite{pfeiffer2024}. The transition at $B^{*}$ is observed for all pressures, demonstrating that phase I is not responsible for the transition between the two SC states~\cite{siddiquee2023,semeniuk2024}. However, the quantum critical fluctuations of its OP may be responsible for the SC pairing mechanism~\cite{nogaki2022,khanenko2025c}, as the AFM order in the cuprates and pnictides sets the $d$- and $s_{\pm}$-wave pairing in these systems, respectively~\cite{monthoux1992,dahm2009}.

Although the phase diagram of \CRA\ in a magnetic field and under pressure is now well established, there is still no consensus on the microscopic nature of the order parameter of phase I, with different proposals involving not just magnetic dipoles, but electric quadrupoles too, having been published. The ordering vector and orientation of the magnetic dipole moment also remain under debate. Based on NQR~\cite{kibune2022} and NMR~\cite{ogata2024} experiments, an antiferromagnetic (AFM) order was proposed, in which opposite magnetic moments polarized along the c-axis develop on the two Ce ions in each unit cell. On the other hand, $\mu$SR experiments indicate a significant in-plane component of the magnetic moment, and a broad field distribution at \To\ suggests incommensurate ordering~\cite{khim2025,dalmas2016}. Furthermore, inelastic neutron scattering experiments have detected quasi-two-dimensional spin fluctuations above \To\ at a commensurate wave vector $(\pi,\pi,0)$~\cite{chen2024} along which ARPES measurements observed substantial nesting~\cite{wu2024}. Theoretical investigations have also produced inconsistent results: calculations based on a localized $4f$-electron CEF model of coupled multipole moments have suggested zero-field easy-plane dipolar order and field-induced quadrupolar order of the $xy$ type~\cite{schmidt2024,thalmeier2025}. This model has been shown to reproduce the phase diagram of phase I in both field directions with remarkable precision. Similarly, two-dimensional FLEX calculations suggest a $(\pi,\pi,0)$ order with in-plane polarization~\cite{nogaki2022}. However, renormalized band structure calculations indicate that in addition to commensurate nesting vectors the Fermi surface also exhibits incommensurate nesting vectors~\cite{hafner2022}. Finally, it has been argued that the nonsymmorphic space group should generically favor incommensurate spin-density-wave (SDW) order~\cite{lee2025}.

Resolving the mystery behind the OP of phase I is crucial for understanding superconductivity in \CRA, given the evidence suggesting that superconductivity emerges at the QCP of phase I. Furthermore, phase I could function as a symmetry-breaking field, lifting the degeneracy of a two-component SC state in a manner analogous to the role played by the AFM order in UPt$_3$~\cite{joynt2002}. The aim of this work is, therefore, to address two crucial questions: (1) Can a two-component SC OP be excluded as an explanation for the multi-phase superconductivity in \CRA, and (2) what is the nature of the OP of phase I? To this end, we have taken measurements of sound velocity in single-crystalline samples of \CRA\ at temperatures as low as 50\,mK and in magnetic fields of up to 16\,T across both transition temperature $T_0\approx0.5\;\rm{K}$ and $T_c\approx0.3\;\rm{K}$. Additional effort was made to take measurements under a hydrostatic pressure of 0.7\,GPa. Our results strongly suggest that the SC order parameter is single-component in both phases and the presence of an incommensurate magnetic order parameter in phase I.

\sout{
}
\*{Speed of sound measurements}

Ultrasound is a powerful method for identifying the symmetry of the order parameters. The velocity of sound-wave propagating along the direction $k_i$ with polarization $p_j$ is related to the elastic constants through $v(\vec{k}_i,\vec{p}_j) =\sqrt{C_{ij} /\rho}$, with the density of the material $\rho$. The propagation direction and the polarization of the sound wave is controlled by using different piezoelectric transducers. The symmetry of the order parameters associated with a phase transition can be revealed by monitoring the change of the elastic constants $C_{ij}$ which informs us of how the strain associated with an elastic constant $C_{ij}$ is coupled to the order parameters and constraints the symmetry of the order parameters. For example, the symmetry of the superconducting order parameters of unconventional superconductors such as Sr$_2$RuO$_4$~\cite{ghosh2021,benhabib2021}, UTe$_2$~\cite{Theuss}, and UPt$_3$~\cite{bishop1984} have been investigated by using ultrasound propagation. 

\begin{figure}[ht!]
	\centering
	\includegraphics[width=1\textwidth]{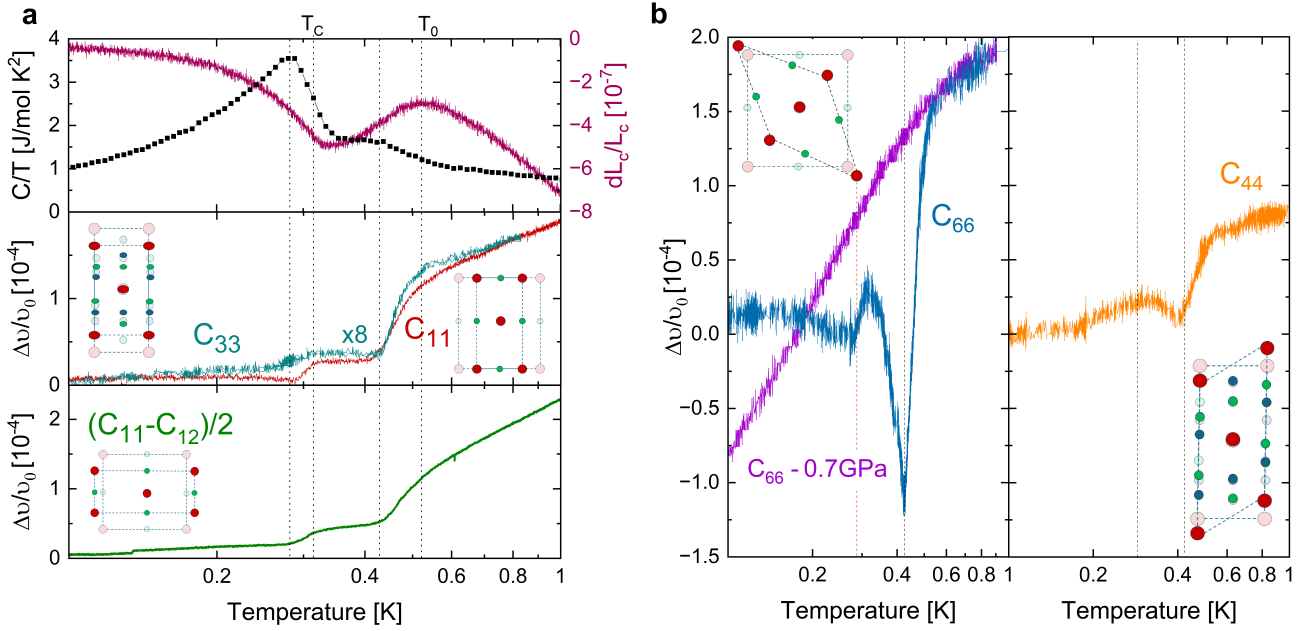}
	\caption{
		\textbf{Temperature dependence of the elastic constants of \CRA\ across the phase transitions.} 
		(a) Comparison of the temperature dependence of the specific heat, $c$-axis thermal expansion and the speed of sound representing the elastic modes  $C_{11}$,  $C_{33}$,  and$(C_{11}-C_{12})/2$  across the zero-field transitions. Dashed lines serve as guides to the eye. (b) Temperature dependence of the speed of sound of the  shear modes $C_{66}$ (blue) and $C_{44}$ (orange). The $C_{66}$ mode exhibits a clear discontinuity at the superconducting transition at ambient pressure. The application of 0.7\,GPa (violet) suppresses both transitions, indicating a single component OP for the SC1 phase. The insets represent the lattice distortion associated with each elastic mode.}
	\label{fig2}
\end{figure}

To investigate the properties of the ordered phases in \CRA\ we have performed measurements of the speed of sound related to the elastic constants $C_{11}$, $C_{33}$, $C_{66}$, $C_{44}$, and $(C_{11}-C_{12})/2$, using the pulsed echo technique at ambient pressure and at a hydrostatic pressure of 0.7\,GPa, down to 50 mK and in applied magnetic fields up to 16\,T. Figure~\ref{fig2} shows the summary of our ultrasonic experiments performed at $B = 0$, additional data in magnetic fields can be found in the Supplementary Material ~\cite{SM}. In Fig.~\ref{fig2}(a) we compare the temperature dependence of the specific heat and $c$-axis thermal expansion (upper panel) with the longitudinal sound-velocity associated with the compressive elastic modes $C_{11}$ and $C_{33}$: both elastic modes exhibit clear step-like features at
the onset of the AFM ordering at \To\ $\approx 0.5$\,K and the superconducting transition at \Tc\ $\approx 0.3$\,K. Although the overall shape of the temperature dependence of $C_{11}$ and $C_{33}$ is very similar, the changes observed in the $C_{33}$ mode are about 8 times smaller than those seen in the $C_{11}$ mode. Slightly less-sharp step-like features with similar magnitude to those seen in $C_{11}$ are also observed in the $(C_{11}-C_{12})/2$ which is related to the sheer mode $\varepsilon_{\rm{B_{1g}}}\equiv\varepsilon_{xx}-\varepsilon_{yy}$ of $B_{1g}$ symmetry of the $D_\mathrm{4h}$ symmetry group.  

More interesting signatures are seen in the remaining shear modes shown in Fig.~\ref{fig2}b. In the case of $C_{66}$, which is associated with the sheer mode $\varepsilon_{xy}$ of $B_{2g}$ symmetry, a pronounced inverse-lambda-like anomaly is observed at $T_0$ suggesting very strong coupling of the $\epsilon_{xy}$ strain to the AFM order parameter. This is consistent with recent thermal expansion experiments which show a tetragonal to orthorhombic distortion at \To~\cite{khanenko2025b}. Moreover, the fact that the  feature observed at \To\ in $C_{66}$ is about 3 times larger than that observed in $(C_{11}-C_{12})/2$ suggests that a large component of the AFM ordered moment should be along the [110] direction. This conclusion, however, ignores the possible role of concomitant quadrupolar density wave order~\cite{hafner2022}, which could also result in a feature at \To. Similar anomalous behavior at $T_0$ appears in $C_{44}$, related to the two-component sheer modes $\varepsilon_{xz}$ and $\varepsilon_{yz}$ of $E_{g}$ symmetry. 

Interestingly, within experimental accuracy, no anomaly is present in the $C_{44}$ elastic constant at $T_c$, which is in stark contrast to the step-like discontinuity seen in $C_{66}$. 
In previous studies of Sr$_2$RuO$_4$, for instance, a discontinuity in the elastic constant $C_{66}$ at $T_c$ has been taken as strong evidence for a two-component SC order parameter~\cite{ghosh2021,benhabib2021}, even though other interpretations are currently considered~\cite{mackenzie2020,willa2021}. In \CRA\, the presence of phase I might induce such a discontinuity, since it can originate from the interaction between the magnetic order and superconductivity, and also the tetragonal to orthorhombic distortion accompanying the magnetic transition. This would explain the behavior of the elastic constants in a magnetic field aligned with the crystallographic $c$-axis (see Fig.~S1 of SM~\cite{SM}). The step-like feature in both $C_{11}$ and $C_{66}$ disappears at $B = 6$\,T, i.e. at a field strong enough to suppress phase I (cf. Fig.~\ref{fig1}). Just a kink is then observed at the SC transition into SC2. This still leaves us with the question of which signature appears in the elastic constants across \Tc\ into phase SC1 in the absence of phase I. Previous transport experiments under pressure\cite{pfeiffer2024} demonstrated that at a modest pressure of 0.5\,GPa phase I is fully suppressed leaving superconductivity intact. Thus to avoid the spurious influence of phase I we have performed additional ultrasound measurements of $C_{66}$ under a hydrostatic pressure of 0.7\,GPa (for experimental details see the methods section). Here we have obtained a smooth curve without any signatures of anomalies at $T_c$ (Fig.~\ref{fig1}b).
\section*{Landau free energy analysis}
\begin{figure}
\centering
\includegraphics[width=0.90\textwidth]{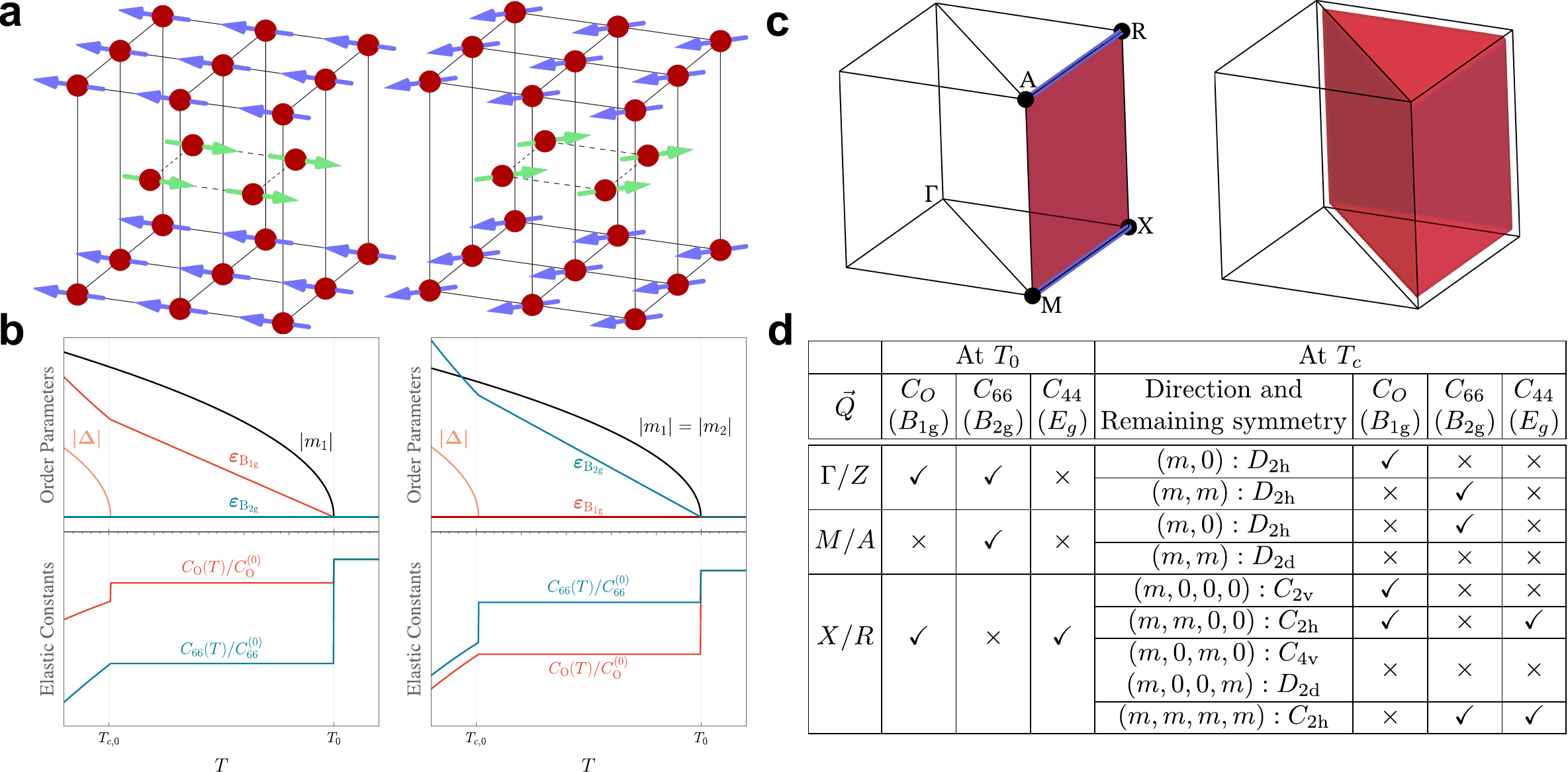}
\caption{
\justifying
\textbf{Summary of theoretical results:} \label{fig:theory}
\textbf{(a)} The two  antiferromagnetic orders with $\vec{Q}=\Gamma(0,0,0)$ corresponding to the $(m,0)$ (left panel) and $(m,m)$ states (right panel). Green and blue arrows represent the local magnetic moments at cerium atoms.
\textbf{(b)} Phase-transition signatures of the order parameters and elastic constants expected from the Landau free energy for the AFM transition with $\vec{Q}=\Gamma(0,0,0)$ and the SC transition. $(\beta_1,\beta_2)=(1,2.5)$ and $(\beta_1,\beta_2)=(1,0.45)$ are used for the left and the right panels, respectively, with the common parameters $(\alpha_0,C^{(0)}_{O},C^{(0)}_{66},\gamma_{1},\gamma_{2})=(1,1,1,0.13,0.2)$.
\textbf{(c)} Ordering vector $\vec{Q}$ compatible with anomalies at the $T=T_0$ transition. Left panel: The high-symmetry lines $Y(k_{x},\pi,0)$ and $T(k_{x},\pi,\pi)$ with $0<k_{x}<\pi$ are marked in blue, and the high-symmetry plane $F(k_{x},\pi,k_{z})$ with $0<k_{x},k_{z}<\pi$ is shown in red. The high-symmetry points (black dots) are not included in these regions. Right panel: The red triangular pillar includes general momentum $(k_{x},k_{x},k_{z})$ with $0<k_{x}<k_{y}<\pi$ and $0<k_{z}<\pi$. Black lines represent high-symmetry lines whose irreducible representations are incompatible with the elastic-moduli anomalies at $T=T_0$. 
(d) Table of elastic constants that can show discontinuous jumps at the magnetic transition at \To\ and at the superconducting transition at \Tc for various time-reversal-invariant ordering vectors $\vec{Q}$ of the magnetic state emerging at \To. The first column in the ``at \Tc'' sector shows the direction of the magnetic order in the representation basis, along with the crystallographic point-groups of the corresponding magnetic states. $B_{1g}$, $B_{2g}$, and $E_{g}$ denote the symmetries of the associated strains. $\checkmark$ indicates allowed strain-order parameter coupling to shear strains, yielding discontinuous elastic jumps; $\times$ indicates cases forbidden by symmetry.
}
\label{fig3}
\end{figure}

To better understand the implications of the observed elastic anomalies for the transition at \To and influence of phase I on the anomalies seen at \Tc, we have employed an analysis based on the phenomenological Landau free energy where the discontinuities in the elastic constants at the phase transitions can be understood as resulting from a magneto-elastic coupling $\varepsilon_{i} M_j M_k$, where $\varepsilon_{i}$ is the strain and $M_i$ the order parameter~\cite{Rehwald1973,benhabib2021,ghosh2021}. Note that the AFM order parameter must be multi-component in order to couple to the shear strains. For brevity, here we consider antiferromagnetic orders with ordering vector $\vec{Q}$ at high-symmetry points in the Brillouin zone, using $\vec{Q}=\Gamma(0,0,0)$ as a concrete example; the analysis for an incommensurate ordering vector is more involved and presented in the Supplementary Material~\cite{SM}. 
Although this is not usually regarded as an AFM ordering vector, in \CRA\ a magnetic state with vanishing moment at this vector is possible, as shown in Fig.~\ref{fig3}(a). Since the AFM order must be at least two-component, the experimental results immediately exclude the single-component $c$-axis polarized $\vec{Q}=\Gamma$ state proposed in~\cite{kibune2022,ogata2024}. The AFM order parameter can be represented as $m_{1}\Gamma_{1}+m_{2}\Gamma_{2}$ where $\Gamma_{1}$ and $\Gamma_{2}$ is the basis of the two-component AFM corresponding to $x$- and $y$-axis-oriented moments, and $m_1$ and $m_2$ are the coefficients. The Landau free energy is then written as
\begin{align}
F =& F_{\rm{m}}+F_{\rm{es}}+F_{\rm{m-es}},\label{eq:F}\\
F_{\rm{m}} =& \alpha_{0}(T-T_{0}) (m_{1}^2+m_{2}^2) + \beta_{1}  (m_{1}^4+m_{2}^4) + \beta_{2}  m_{1}^{2}m_{2}^2,\label{eq:Fm}\\
F_{\rm{es}} =& \frac{1}{2} C^{(0)}_{A}\varepsilon_{\rm{A_{1g}}}^{2}
 + \frac{1}{2} C^{(0)}_{O} \varepsilon_{\rm{B_{1g}}}^{2}
 + \frac{1}{2} C^{(0)}_{66} \varepsilon_{\rm{B_{2g}}}^{2}, 
 \label{eq:Fes}\\
 F_{\rm{m-es}}=& \gamma_{1} {\epsilon_{\rm{B_{1g}}}} (m_{1}^{2}-m_{2}^{2}) + \gamma_{2} {\epsilon_{\rm{B_{2g}}}} m_{1}m_{2}+\gamma_{3} {\epsilon_{\rm{A_{1g}}}}(m_{1}^2+m_{2}^2),\label{eq:Fmes}
\end{align}
where $F_m$, $F_{es}$, and $F_{\rm{m-es}} $ are the free energies for the magnetic order, elastic strain, and the magneto-elastic coupling, respectively. Here $C^{(0)}_{A}\equiv (C^{(0)}_{11}+C^{(0)}_{12})/2$, $C^{(0)}_{O}\equiv (C^{(0)}_{11}-C^{(0)}_{12})/2$, and $C^{(0)}_{66}$ correspond to the \emph{bare} elastic moduli defined at temperatures above the transition. We do not include bi-quadratic couplings since they do not provide any constraints on the symmetry of the order parameter.
Furthermore, since the magneto-elastic coupling to an $A_{1g}$ strain ${\epsilon_{\rm{A_{1g}}}}$ is always allowed by symmetry, it is not crucial in deciding the symmetry of the order parameters, and we henceforth ignore this coupling without affecting the final conclusion.

Depending on $\beta_{1}$ and $\beta_{2}$, the magnetic Landau free energy $F_m$ in Eq.~\ref{eq:F} yields two kinds of AFM ground states: $(m_1,m_2)=(m,0)$ for $2\beta_1<\beta$, and $(m_1,m_2)=(m,m)$ for $2\beta_1>\beta_2$, which are shown in Fig.~\ref{fig3}(a). Both cases result in an orthorhombic magnetic phase, although with different strain profile: for  $(m_1,m_2)\propto(m,0)$ the two-fold rotation axes along the tetragonal crystal axes are preserved, which allows a finite $\varepsilon_{\rm{B_{1g}}}$ strain since it transforms trivially under these symmetry operations. On the other hand, in the $(m_1,m_2)\propto(m,m)$ state the in-plane axes of the remaining two-fold rotations are aligned along the $\hat{x}\pm\hat{y}$ directions of the tetragonal system. Consequently, it is the $B_{\rm 2g}$ irreducible representation of the tetragonal system that transforms trivially in this magnetic phase and $\varepsilon_{\rm{B_{2g}}}$ can be finite after the phase transition. Notably, recent thermal expansion experiments evidence a tetragonal-to-orthorhombic lattice distortion at \To~\cite{khanenko2025b}.
%

The development of the order parameters and strains is reflected in discontinuities of the  elastic moduli $C_{O}(T)$ and $C_{66}(T)$ upon crossing \To, as shown in Fig.~\ref{fig:theory}. To obtain these results we take the second derivative of the minimum of the Landau free energy with respect to the strain fields~\cite{Rehwald1973} (See {\emph{Method}} section). Notably, although the symmetry groups of the two AFM phases are different, it is not possible to distinguish between the two phases based on the behaviour of the elastic moduli at the transition. 
The symmetry-lowering due to the AFM order nevertheless affects the elastic response at the superconducting transition. To see this, we extend Eq.~\ref{eq:F} by including the SC OP $\Delta$, i.e. $F\rightarrow F + F_{\Delta}$ with
\begin{equation}
    F_{\Delta}= \alpha_{{\rm sc}}(T-T_{{\rm c,0}})\Delta^{2}+\beta_{{\rm sc}}\Delta^{4} + \zeta_{1}\varepsilon_{{\rm B}_{1g}}(m_{1}^{2}-m_{2}^{2})\Delta^{2}+\zeta_{2}\varepsilon_{{\rm B}_{2g}}m_{1}m_{2}\Delta^{2}
\end{equation}
where $T_{c,0}$ is the bare superconducting transition. We ignore the direct coupling between the AFM and SC in order to focus on the how the AFM order affects the coupling to the strain field, which is included in the final two terms. 
Although for a tetragonal system a discontinuity in $C_{O}$ or $C_{66}$ should only be present if the SC OP is multi-component, if $\varepsilon_{\rm{B_{1g}}}$ or  $\varepsilon_{\rm{B_{2g}}}$ is finite at $T<T_0$, the strain-order parameter coupling which is linear in the shear strain and quadratic in superconducting order parameter of arbitrary symmetry is allowed. As such, a  discontinuous jump of $C_{O}$ at the superconducting transition is expected for the $(m,0)$ AFM order, whereas a discontinuous jump of $C_{66}$ occurs in the presence of the $(m,m)$ AFM order.

The $\vec{Q}=\Gamma$ example discussed above shows how the elastic response at both \To\ and \Tc\ conveys information about the nature of the magnetic state. In the table presented in Fig.~\ref{fig3}d we summarize predictions for the behavior of the elastic moduli in all cases with an ordering vector at a time-reversal invariant momentum $\vec{Q}$, i.e. when $\vec{Q}$ is identical to $-\vec{Q}$ up to a reciprocal lattice vector. 
A detailed analysis for all other ordering vectors is presented in the supplemental material~\cite{SM}. 
Comparing the predictions for discontinuities for time-reversal invariant $\vec{Q}$ with the results of our ambient pressure experiments in \CRA\ reveals that none of the AFM states can account for the observed discontinuities: in particular, in no case are discontinuities observed in all shear elastic constants at \To, or in both $C_{O}$ and $C_{66}$ at \Tc.

The conclusion of our Landau analysis is that the observed behaviour of the elastic constants can only be explained by an AFM ordering vector located away from the time-reversal invariant momentum. Moreover, requiring the couplings linear that are linear in a shear strain and quadratic in the AFM order parameters rules out the location of the ordering vector on the majority of high-symmetry lines and  planes in the Brillouin zone. The allowed ordering vector thus must be located at $Y(k_x, \pi, 0)/T(k_x, \pi, \pi)$, $F(k_x, \pi, k_z)$, and general $(k_x, k_y, k_z)$ with $k_x \neq k_y$ where all the unspecified momentum variables are neither $0$ or $\pi$. These regions correspond to the regions colored in red in Fig.~\ref{fig:theory}b.  In {\emph{Methods}}, we present the Landau free energy with magneto-elastic couplings for the case of $\vec{Q}=Y(k_x,\pi,0)$, which involves the minimal number of order parameters among the three cases. The ground state solution relevant to our ultrasound results is discussed in more detail in the supplemental material~\cite{SM}.

Finally, we turn to the issue of whether the SC order parameter is multicomponent. Due to the apparent lowering of the lattice symmetry from tetragonal to orthorhombic at \To~\cite{khanenko2025b}, the observed discontinuities in $C_{66}$ and $(C_{11}-C_{12})/2$ at \Tc\ are not reliable indicators of a multi-component gap. In this work we have performed ultrasonic experiments in two very distinct points of the phase diagram: At ambient pressure (where SC1 coexists with phase I) and at 0.7\,GPa (where phase I is fully suppressed~\cite{pfeiffer2024}). In the latter case the $C_{66}$ elastic constant shows no feature at \Tc\, which is consistent with the jump at ambient pressure being caused by the presence of phase I, and not reflecting an intrinsic property of the SC OP. This strongly supports a single-component SC OP,
similarly to the case of  UTe$_2$ \cite{Theuss}.
\section*{Discussion}
The analysis presented here relies on the presence of fully formed long range order (LRO) with infinite correlation length $\xi_{m}$. It is was however shown in other heavy fermion systems that close to a quantum critical point the full LRO may not be developed. For example in UPt$_{3}$ $\xi_{m}/a \approx 45$ (with $a$ being the in-plane lattice constant). In such a case the magnetic state should rather be seen as a state where the orientation of moments varies spatially with a typical length scale of $\xi_{m}$. If this would also be the case for \CRA\ then phase I would be best described by a superposition of the order parameters in Fig.~\ref{fig:theory}a with spatially varying amplitude. In that case for  $\xi_{m} \ll \lambda$, where $\lambda$ is the wavelength of  ultrasound ($\lambda \approx 10-100\rm um$), any sound mode could in principle couple to the spatially modulated magnetic order, thus invalidating the derived selection rules. The most direct way to exclude such scenario would be to directly measure $\xi_{m}$ using INS. However, to date INS experiments are still inconclusive, possibly due to the small ordered moment in \CRA~\cite{chen2024}. 
Another concern is that in strongly correlated electron systems - and in particular in HF systems like \CRA\ - ultrasonic evidence can be difficult to interpret. A typical example is UPt$_{3}$ in which the SC order parameter is known to be multi-component, but unexpectedly no discontinuity was observed in the elastic constant C$_{66}$ at \Tc~\cite{bruls1990,bruls1991}. Likewise, in Sr$_2$RuO$_4$, a clear discontinuity in $C_{66}$ was observed at \Tc~\cite{ghosh2021,benhabib2021}, but no discontinuity has been observed in $(C_{11}-C_{12})/2$, which is inconsistent with a SC OP belonging to a single IR~\cite{willa2021}, and contradicts several other experiments which point to a single-component SC OP~\cite{mackenzie2020}. In contrast to these cases, however, our results are entirely consistent with a straightforward theoretical model.

In summary, our ultrasound measurements and symmetry analysis reveal that phase I in \CRA\ most likely cannot arise from any magnetic ordering vector at the high-symmetry points of the Brillouin zone. Instead, the analysis of the experiments points toward an incommensurate antiferromagnetic state. By extending the measurements to high magnetic fields and to a pressure of 0.7\,GPa, at which phase I is fully suppressed, we isolated the superconducting response and show that both SC phases host a single-component order parameter. Together, these results impose stringent new constraints on the microscopic order in \CRA\ and highlight its unique position among heavy-fermion systems.

\newpage
\section*{Methods}
\subsection*{Sample preparation and ultrasound propagation measurements} 

\CRA\textcolor{white}{a}single crystals were grown in Bi flux as described in Ref. \cite{khim2021}. The
composition and the crystal structure were determined by energy-dispersive x-ray spectroscopy (EDXS) and single crystal x-ray diffraction analysis.

In the speed of sound experiments we measured the relative change of the sound velocity, $dv/v$ using a phase-sensitive pulse-echo technique. Two piezoelectric lithium niobate (LiNbO\textsubscript{3}) resonance transducers were glued to opposite parallel surfaces of the sample to excite and detect acoustic waves. Selected samples of ~0.5-1 mm in diameter were oriented using Laue diffraction for precise alignment of magnetic fields and sound propagation directions with crystallographic axes. Subsequently, the sample surfaces were polished using a focused ion beam milling in order to ensure that the transducers were attached to smooth and parallel surfaces. In contrast to traditional polishing methods ion milling ensured the samples were not mechanically strained, helping to avoid deformations and internal cracking during polishing. We have previously successfully employed this method for the preparation of very delicate layered Van der Waals materials for ultrasonic experiments \cite{galeski2021origin, Baptiste}. For experiments under pressure we have used a commercially available pressure cell optimized for ultrasound experiments from C\&T Factory Co.,Ltd and used the Daphne 7474  oil as the pressure medium. Due to the fragility of the ultrasound installation for those experiment we have used crystals from a newer \CRA batch with bigger crystals ($l= 1.53mm$) and slightly higher transition temperatures. To determine the applied pressure we have used a pure tin wire and determined the pressure by measuring the shift of its superconducting transition\cite{PressureTin}, measurements of the R(T) of the tin wire were carried out in the PPMS taking care that there would be no trapped magnetic flux in the magnet. All ultrasound measurements were performed in a Kelvinox 400 dilution refrigerator from Oxford Instruments. 

\subsection*{Specific heat and thermal expansion}
The specific heat was measured using a semi-adiabatic compensated heat-pulse method described in Ref.~\cite{wilhelm2004}. The nuclear contribution due to the spin $I = 3/2$ As nuclei is negligible at zero field lowest to the lowest  temperature measured (see SM of Ref.~\cite{semeniuk2023}). High-resolution thermal expansion was measured by means of a capacitive CuBe dilatometer~\cite{kuechler2012}. All low-temperature measurements were performed in $^{3}$He/$^{4}$He dilution refrigerators.

\subsection*{Analytical expression for the ground state solution of the Landau free energy and the discontinuity of the elastic constants at the phase transition.}

In the main text, we have presented the characterizing feature of the ground state solution of the Landau free energy for the antiferromagnetic state with the ordering vector $\vec{Q}=\Gamma(0,0,0)$. Here, we provide the analytical expressions for those solutions.
\begin{align}
2\beta_1<\beta_2:&\quad (m_1,m_2)=
\sqrt{\frac{-\alpha}{2\beta_1-\beta_{\rm{B_{1g}}}}}(1,0),\;
&&(\varepsilon_{\rm{B_{1g}}},\varepsilon_{\rm{B_{2g}}})=(\frac{\beta_{\rm{B_{1g}}}}{2\beta_{1}-\beta_{\rm{B_{1g}}}}\frac{\alpha}{\gamma_1},0),\\
2\beta_1>\beta_2:&\quad (m_1,m_2)=
\sqrt{\frac{-\alpha}{2\beta_1+\beta_2-\beta_{\rm{B_{2g}}}}}(1,1),
\;&&(\varepsilon_{\rm{B_{1g}}},\varepsilon_{\rm{B_{2g}}})=(0,\frac{2\beta_{\rm{B_{2g}}}}{2\beta_1+\beta_2-2\beta_{\rm{B_{2g}}}}\frac{\alpha}{\gamma_2}),
\end{align}
with $\beta_{O}=\gamma_1^2/C^{(0)}_{O}$, $\beta_{\rm{B_{2g}}}=\gamma_2^2/C^{(0)}_{66}$, and $\alpha=\alpha_0 (T-T_0)$.

The change of the elastic moduli $C_k$ from the bare value are obtained through $C_{k}-C^{(0)}_{k}=-\sum_{i,j}\frac{\partial^{2}F}{\partial \varepsilon_k \partial m_i }[\mathbb{M}^{-1}]_{i,j}\frac{\partial^{2}F}{\partial m_j \partial \varepsilon_k}$ with $\mathbb{M}_{i,j}=\frac{\partial^{2}F}{\partial{m_i}\partial{m_j}}$. This is the second derivative of the minimum of the Landau free energy with respect to the strain fields. The discontinuities of elastic moduli for the solutions given above are written as
\begin{align}
2\beta_1<\beta_2:&\quad \frac{C_{O}-C^{(0)}_{O}}{C^{(0)}_{O}}=-\frac{\beta_{\rm{B_{1g}}}}{2\beta_1},&& \;\frac{C_{66}-C^{(0)}_{66}}{C^{(0)}_{66}}=-\frac{2\beta_{\rm{B_{2g}}}}{-2\beta_1+\beta_2+2\beta_{\rm{B_{1g}}}},\\
2\beta_1>\beta_2:&\quad \frac{C_{O}-C^{(0)}_{O}}{C^{(0)}_{O}}=-\frac{2\beta_{\rm{B_{1g}}}}{2\beta_{1}-\beta_{2}+2\beta_{\rm{B_{2g}}}},&& \;\frac{C_{66}-C^{(0)}_{66}}{C^{(0)}_{66}}=-\frac{2\beta_{\rm{B_{2g}}}}{2\beta_1+\beta_2}.
\end{align}
\subsection*{Phenomenological analysis of the phase transition into the incommensurate ordering with $\vec{Q}=Y(k_x,\pi,0)$}
In the main text, we discuss the relevance of incommensurate ordering with ordering vector $\vec{Q}$ on the high-symmetry line $Y(k_x,\pi,0)$, in the high-symmetry plane $F(k_x,\pi,k_z)$, or at a general point ${\rm{GP}}(k_x,k_y,k_z)$ as a candidate for the phase transition at $T=T_0$. Each case involves eight, sixteen, and sixteen order parameters, respectively, and we present the Landau free energy for the order parameter with $\vec{Q}$ on the line $Y$ for the simplicity as there is no fundamental difference when it comes to the symmetry analysis for the linear-quadratic coupling. It is given by 
\begin{align}
F_{Y}=&F_{m}+F_{me}+\frac{1}{2} C^{(0)}_{\rm{B_{1g}}}{\varepsilon^{2}_{\rm{B_1g}}}
 + \frac{1}{2} C^{(0)}_{\rm{B_{2g}}} {\varepsilon^{2}_{\rm{B_{2g}}}}
 + \frac{1}{2} C^{(0)}_{\rm{E_{g}}} (\varepsilon_{xz}^{2}+\epsilon_{yz}^{2}),\\
F_{m}=&\alpha I_{\rm{A_{1g}}} +\beta_1 I_{\rm{A_{1g}}}^2 + \beta_2 I_{\rm{B_{1g}}}^2+\beta_3 I_{\rm{A_{2u}}}^2+\beta_4 I_{\rm{B_{2g}}}^2+
\beta_5 I_{\rm{B_{2u}}}^2 + \beta_6 (I_{xz}^2+I_{yz}^2),\\
F_{me}=&\gamma_1 \varepsilon_{\rm B_{1g}}I_{\rm{B_{1g}}}+\gamma_2 \varepsilon_{\rm B_{2g}}I_{\rm{B_{2g}}}+\gamma_3 (\varepsilon_{xz}I_{xz}+\varepsilon_{yz}I_{yz})
\end{align}
with $\alpha=\alpha_0(T-T_0)$. Here, $I_{\rm{A_{1g}}}=\sum_{i=1,..,8}m_i^2$, $I_{\rm{B_{1g}}}=\sum_{i=1,..,4}m_i^2-\sum_{i=5,..,8}m_i^2$,  $I_{\rm{B_{2g}}}=m_1 m_4+m_3 m_4 -m_5 m_6 -m_7 m_8$, $I_{xz}=m_1^{2} -m_2^{2}+m_3^{2}-m_4^{2}$, $I_{yz}=m_5^{2} -m_6^{2}+m_7^{2}-m_8^{2}$, and other $I_\alpha$'s are defined in accordance with the symmetry indicated by the label (See supplemental material for detail.). The case, which yields finite $\varepsilon_{\rm B_{1g}}$ and $\varepsilon_{\rm B_{2g}}$ at $T<T_0$ with associated discontinuity of $C_{\rm B_{1g}}$ and $C_{\rm B_{2g}}$ while $\varepsilon_{xz}=\varepsilon_{yz}=0$ occurs when $I_{\rm{B_{1g}}},I_{\rm{B_{2g}}}\neq 0$, and $I_{xz}=I_{yz}=0$ hold in $T<T_0$. For example, $m_1=m_2=m_3=m_4\neq 0$ and $m_5=m_6=m_7=m_8=0$ is a case, which also gives rise to a discontinuity of the elastic moduli $C_{xz}$ with respect to $\varepsilon_{xz}$ at $T=T_0$ which is identical to $C^{(0)}_{\rm E_{g}}$ at $T>T_0$. Therefore, this is the case consistent with the anomalies observed in the ultrasound result.

\bibliography{main_final_manuel.bib}

\subsection*{\textbf{Data availability}}

All data generated or analyzed during this study are included in this published article and its Supplementary Material. 

\section*{Acknowledgements (not compulsory)}
We are indebted to C. Geibel, J. F. Landaeta, K. Semeniuk, O. Stockert, S. Wiedmann and G. Zwicknagl, for insightful discussions. We acknowledge support from the DFG through the Würzburg-Dresden Cluster of Excellence on Complexity, Topology and Dynamics in Quantum Matter – \textit{ctd.qmat} (EXC 2147, project-id 39085490), the Collaborative Research Center SFB 1143 (Project No. 247310070) and by Hochfeld-Magnetlabor Dresden (HLD) at HZDR and HFML-FELIX members of the European Magnetic Field Laboratory (EMFL). CL and PMRB were supported by the Marsden Fund
Council from Government funding, managed by Royal
Society Te Ap\={a}rangi, Contract No. UOO2222.

\section*{Author contributions statement}

S.G and M.B. conceived the experiments. S.G., F.B., A.T.M.B, R.A., J.S., Z.S. and J.W. conducted the  ultrasonic experiments. P.K., and M.B. conducted specific heat and thermal expansion measurements. S.K. grew and prepared the \CRA crystals. S.G., C.L., P.T., P.M.R.B., M.B. analized the experimental results. E.H was involved in the interpretation of the results. C.L. and P.M.R.B developed the theoretical model. S.G., C.L., P.M.R.B and M.B. wrote the manuscript. All authors reviewed the manuscript.

\subsection{Ethics declarations}

\subsubsection{Competing interests}

The authors declare no competing interests.

\end{document}